\documentclass[prb,twocolumn,superscriptaddress,showpacs]{revtex4}
\usepackage[dvips]{graphicx}
\usepackage{latexsym}
\usepackage{bm}
\begin{document}
\newcommand{\fig}[2]{\includegraphics[width=#1]{#2}}

\newcommand{\dprime}{{\prime\prime}}
\newcommand{\be}{\begin{equation}}
\newcommand{\den}{\overline{n}} 
\newcommand{\ee}{\end{equation}}
\newcommand{\bea}{\begin{eqnarray}} 
\newcommand{\eea}{\end{eqnarray}}
\newcommand{\nn}{\nonumber} 
\newcommand{\bk}{{\bf k}}
\newcommand{\vN}{{\bf \nabla}}
\newcommand{\vA}{{\bf A}}
\newcommand{\vE}{{\bf E}}
\newcommand{\vj}{{\bf j}}
\newcommand{\vJ}{{\bf J}}
\newcommand{\bs}{{\bf S}}
\newcommand{\vn}{{\bf v}_n}
\newcommand{\vv}{{\bf v}} 
\newcommand{\la}{\langle}
\newcommand{\ra}{\rangle} 
\newcommand{\ph}{\phi} 
\newcommand{\dg}{\dagger}
\newcommand{\br}{{\bf{r}}} 
\newcommand{\bo}{{\bf{0}}} 
\newcommand{\bR}{{\bf{R}}} 
\newcommand{\bS}{{\bf{S}}} 
\newcommand{\bq}{{\bf{q}}}
\newcommand{\bQ}{{\bf{Q}}}
\newcommand{\vQ}{{\bf{Q}}} 
\newcommand{\hj}{\hat{\alpha}}
\newcommand{\hx}{\hat{\bf x}} 
\newcommand{\hy}{\hat{\bf y}}
\newcommand{\hz}{\hat{\bf z}}
\newcommand{\vS}{{\bf S}} 
\newcommand{\cV}{{\cal U}}
\newcommand{\cD}{{\cal D}} 
\newcommand{\tnh}{{\rm tanh}}
\newcommand{\sh}{{\rm sech}} 
\newcommand{\vR}{{\bf R}}
\newcommand{\crx}{c^\dg(\vr)c(\vr+\hx)}
\newcommand{\crkubox}{c^\dg(\vr)c(\vr+\hat{x})}
\newcommand{\pll}{\parallel} 
\newcommand{\crj}{c^\dg(\vr)c(\vr+\hj)}
\newcommand{\crmj}{c^\dg(\vr)c(\vr - \hj)}
\newcommand{\sumall}{\sum_{\vr}} 
\newcommand{\sumx}{\sum_{r_1}}
\newcommand{\nabj}{\nabla_\alpha \theta(\vr)} 
\newcommand{\nabx}{\nabla_1\theta(\vr)} 
\newcommand{\sumy}{\sum_{r_2,\ldots,r_d}}
\newcommand{\krj}{K(\vr,\vr+\hj)} 
\newcommand{\sigr}{|\psi_0\rangle}
\newcommand{\sigl}{\langle\psi_0 |}
\newcommand{\sier}{|\psi_{\Phi}\rangle}
\newcommand{\siel}{\langle\psi_{\Phi}|}
\newcommand{\sumrj}{\sum_{\vr,\alpha=1\ldots d}}
\newcommand{\krw}{K(\vr,\vr+\hx)} 
\newcommand{\Dtheta}{\Delta\theta}
\newcommand{\rhonew}{\hat{\rho}(\Phi)}
\newcommand{\rhoold}{\hat{\rho_0}(\Phi)} 
\newcommand{\dt}{\delta\tau}
\newcommand{\cP}{{\cal P}} 
\newcommand{\cS}{{\cal S}}
\newcommand{\vm}{{\bf m}} 
\newcommand{\hnr}{\hat{n}({\vr})}
\newcommand{\hnm}{\hat{n}({\vm})} 
\newcommand{\del}{\hat{\delta}}
\newcommand{\upa}{\uparrow} 
\newcommand{\dna}{\downarrow}
\newcommand{\dnk}{\delta n_{\vk}}
\newcommand{\dnks}{\delta n_{\vk,\sigma}}
\newcommand{\dnkp}{\delta n_{\vk '}}

\title{$XY$ ring-exchange model on the triangular lattice}
\author{Leon Balents}
\affiliation{Department of  Physics, University of California,
Santa Barbara, CA 93106--4030}
\author{Arun Paramekanti}
\affiliation{Department of  Physics, University of California,
Santa Barbara, CA 93106--4030}
\affiliation{Kavli Institute for Theoretical Physics,
University of California, Santa Barbara, CA 93106--4030}
%\author{Claudia Peca??}
%\affiliation{Department of  Physics, University of California,
%Santa Barbara, CA 93106--4030}
\begin{abstract}
%\vspace{0.1cm}
  We study ring-exchange models for bosons or $XY$-spins on
  the triangular lattice. A four-spin exchange leads to a manifold of
  ground states with gapless excitations and critical power-law
  correlations.  With a nearest-neighbour exchange, fluctuations
  select a four-fold ferrimagnetically ordered ground state with a
  small spin/superfluid stiffness which breaks the global $U(1)$ and
  translational symmetry.  We explore consequences for phase
  transitions at finite temperature and in an in-plane magnetic field.
  \typeout{polish abstract}
\end{abstract}

\maketitle

\section{introduction}

Multi-spin exchange models incorporating ring-exchange processes have
been of interest since the early studies of magnetism in solid
Helium-3 \cite{thouless,hetherington}.  Ring-exchange processes could
also play an important role in Wigner crystals near the melting
density \cite{ceperley, chakravarty}, and in Mott insulators which
retain a fair degree of local charge fluctuations \cite{macdonald}.
Indeed, neutron scattering experiments \cite{coldea} in 
insulating La$_2$CuO$_4$ have shown that aspects of the spin wave dispersion 
in the antiferromagnet may be understood by invoking ring-exchange terms. 
Another reason for the interest in such models is that they may support
spin-liquid phases which are translationally invariant Mott insulators
with no magnetic order, as indicated from numerics on triangular and
Kagome lattices \cite{lhuillier}.
%
%Early mean-field studies of ring-exchange models found a variety
%of ordered states {chubukov, lhuillier}. Around the
%same time, variational studies {kubo} showed that one might
%obtain
%
This has been established analytically in some models with U(1)
symmetry \cite{leon,motrunich}, which may also be viewed as boson models.
Finally, many models such as the
quantum dimer model on a triangular lattice \cite{moessner}, the
easy-axis version of a generalized Heisenberg model on the Kagome
lattice \cite{leon} and the easy-axis antiferromagnet on the
pyrochlore lattice \cite{hermele} may be mapped onto effective
ring-exchange models in their low energy subspace of states. Thus,
understanding the phases and phase transitions in ring exchange models
is important.

In this paper, we will focus on an $XY$ (``easy-plane'') ring-exchange model 
on the triangular lattice with both 4-spin and 2-spin
exchange terms (with strengths $K$ and $J$ respectively) in the regime
$J\!\!\ll\!\! K$. This model is interesting from several points of
view.  First, several systems such as Wigner crystals and certain
organic Mott insulators form a triangular lattice of spins with
possibly appreciable ring-exchange processes, and the $XY$ spin
problem is one tractable limit of such models.  Second, easy-plane
magnets on triangular lattices are known to exist (though mostly as
stacked layers forming a three dimensional system) and this could be
of some relevance to them --- in fact, these systems have motivated
several studies on easy-plane triangular quantum Heisenberg models
\cite{luca}. Finally, earlier work by some of us has shown that
related models on the Kagome lattice are fractionalized \cite{leon},
whereas they support a critical phase (the ``exciton Bose liquid'') on
the square lattice \cite{bosemet}.  This was essentially shown by
perturbing around the pure ring-exchange limit ($J=0$) where the
Kagome and square lattices respectively have ${\cal O}(L^2)$ and
${\cal O}(L)$ conserved quantities. It is thus worthwhile to examine
cases where such extensive symmetries are absent from the outset even
in the pure ring-exchange model, as is the case on the triangular
lattice.

The principal results of this paper are: (i) We show that the U(1)
four-spin exchange model on the triangular lattice (with $S_z=0$ or
half-filling for the bosons) has a manifold of degenerate ground
states and the correlation functions are critical in the ground
states. We identify the appropriate symmetries and conserved
quantities which give rise to this.  (ii) Perturbing in the nearest
neighbor exchange (boson hopping) we find that fluctuations select a
four-fold set of states from the degenerate manifold and the system
develops long-range order at zero temperature. The ordered states
break the global U(1) symmetry as well as translational symmetry.
(iii) In the ordered ground states, the superfluid/spin stiffness
$\sim J^2/K$ and is very small for small $J/K$.  At any small nonzero
temperature, the U(1) symmetry is restored and we are in a phase with
power-law phase/spin order, which gives way to a disordered phase via
a Berezinskii-Kosterlitz-Thouless \cite{bkt} (BKT) transition for $T >
T_{BKT} \sim J^2/K$. However, the discrete symmetry of the broken
translations is not destroyed until a higher temperature $T_c \sim J$.
We analyze phase transitions at finite temperature and with in-plane
magnetic fields within a Landau theory, and discuss the phase diagram
and experimental consequences.

The outline of the paper is as follows. In Section II, we will present
the Hamiltonian. In Section III we analyze its symmetries and 
conserved quantities,
and define the dual representation of the model which allows us to
analyze the instabilities of the system towards charge and energy
ordering. We present arguments and numerical results to
show that the model with $J=0$ has critical power-law
correlations at $T=0$. In Section IV, we perturbatively analyze
effects on introducing the nearest neighbor exchange (boson
hopping) and find a phase-ordered state. 
We also discuss the phase diagram within Landau theory as a
function of temperature and in-plane magnetic-field. We close with a
discussion and speculations for three-dimensional generalizations and
SU(2) invariant models in Section V.

\section{model}

We define the model in rotor variables as
\bea
H&=&K \sum_P \cos(\varphi_1-\varphi_2+\varphi_3-\varphi_4) \nn\\
&+&J \sum_{\la i,j\ra} \cos(\varphi_i-\varphi_j)
+ \frac{U}{2}\sum_i (n_i-\bar{n})^2
\label{hamiltonian}
\eea
where $\varphi_i$ is the phase of the boson variable ($\varphi_i=0$ and
$\varphi_i=2\pi$ are identified) and $n_i$ is the the canonically
conjugate boson number (respectively the angle and angular momentum of
a $U(1)$ rotor), satisfying the commutation relation
$\big[\varphi_i,n_j\big]=i\delta_{i,j}$. The terms $\varphi_1\ldots\varphi_4$
in the first term of $H$ denote angles around 4-site plaquettes of the
triangular lattice --- there are three such kinds of plaquettes as
shown in Fig.~1(a). The $J$-term denotes nearest-neighbor hopping of
bosons while $U$ denotes a repulsive interaction between bosons. With
$U/K \to \infty$, we can identify $S_z(\br)=n_\br-1/2$,
$S^\pm(\br)=\exp(\pm i\varphi_\br)$ and this model reduces to a $S=1/2$
$XY$ quantum spin model with $\bar{n}=1/2$ corresponding to
total $S_z=0$.

\begin{figure}
\begin{center}
\vskip-2mm
\hspace*{0mm}
\fig{3.0in}{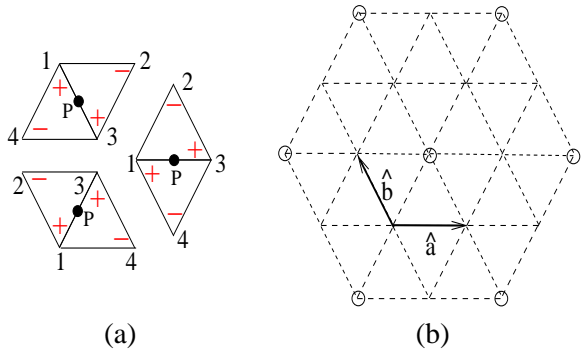}
\vskip-2mm
\caption{(a) The three kinds of plaquettes on the triangular lattice.
The ring-exchange process involves the spins located at the points 1 -- 4 
of each plaquette. The labelling is chosen to coincide with that
appropriate for the definition dual variables in the dual model discussed 
in Section II(C), the sites $P$ form the sites of the dual Kagome lattice. 
(b) Rotating $\varphi \to \varphi+\pi$ on the
sites indicated by open circles (which form one of four possible triangular 
sublattices)
changes the sign of the ring-exchange term $K$ in the Hamiltonian.
Also shown are the basis vectors $\hat{a},\hat{b}$ for the triangular 
lattice.}
\label{fig:plaquette}
\end{center}
\end{figure}

With $J=0$, changing $\varphi_\br\to\phi_\br=\varphi_\br+\pi$ on the sites 
indicated by open circles in Fig.~1(b) changes the sign of $K$ 
(there are four such choices
of a triangular sublattice on which to make this transformation, as is clear
from the figure). Thus, for $J=0$, the ground 
state energy is independent of the sign of $K$.  For nonzero $J$ however,
the sign of $K$ is important.

For $K < 0$, the leading perturbative corrections on including a
nonzero $J$ are ${\cal O}(J)$, and the energy and phases depend on the
sign of $J$ --- for $J<0$ we get a ferromagnetic phase, while $J>0$
leads to a $\sqrt{3}\times\sqrt{3}$ Neel order, which are also the
ground states for large $|J/K|$ (where we may ignore the 4-spin
term). Thus there are no phase transitions at any nonzero $J$ and we
only obtain the well-studied phases.

For $K>0$, as appropriate for say spin degrees of freedom in an
electronic Mott insulator, it turns out that the leading
perturbative corrections to the free energy are ${\cal O}(J^2)$. To
this order, the free energy is independent of the sign of $J$ and we
obtain a ferrimagnetic phase which breaks the global U(1) invariance
as well as four-fold translational symmetry. The physics of this phase
and the phase transitions out of it will be the focus of this
paper. For large $|J|/K$ we of course recover the conventional phases
mentioned above.

To make an estimate of the coupling constants in one case, let us
imagine starting from the Hubbard model for electrons (with nearest
neighbor hopping) at half-filling on a triangular lattice and
perturbing in $t/U$ to derive an effective spin model with
ring-exchange terms in the insulator. This takes the form
\bea
H_{\rm spin}&=&
K \sum_{\Box} 
\left[(\bs_1\cdot \bs_2) (\bs_3\cdot \bs_4)
+(\bs_1\cdot \bs_4) (\bs_2\cdot \bs_3) \right. \nn\\
&-& \left. (\bs_1\cdot \bs_3) (\bs_2\cdot \bs_4)\right]
+ \sum_{i,j} J_{i,j} \bs_i\cdot \bs_j 
\label{hubbard}
\eea
where the first term involves all 4-site plaquettes on the triangular
lattice. Adapting results from Ref.~\cite{macdonald} to this case, we find
$J=4 t^2/U-28 t^4/U^3$, $J'=J''=4 t^4/U^3$, $K=80 t^4/U^3$. For
$U/t=6$, numerical results \cite{imada} show that the model is in an insulating
phase though still close to the metal-insulator boundary; in this case
$J/t=0.53$, $J'/t=J''/t=0.02$, $K/t=0.37$. Clearly, it seems that one
can describe the spin degrees of freedom in the insulator by setting
the further-neighbor couplings $J'=J''=0$ and retaining only nonzero $J,K$.  
Indeed, exact-diagonalization studies \cite{lhuillier} find a spin-liquid phase 
for a closely related model in qualitative agreement with the Monte Carlo results on 
the Hubbard model \cite{imada}. It thus seems profitable to understand the above model 
for $J/K\!\ll\! 1$ as a starting point to analyze this full problem. For
$U/K \to \infty$, the Hamiltonian in Eq.~\ref{hamiltonian} is precisely the $XY$ limit 
of the above model obtained by setting terms containing $S_z$ to zero.

\section{Model with $J=0$}

\subsection{Symmetries}

For $J=0$, we can change the sign of $K$ by shifting $\varphi_\br\to
\phi_\br=\varphi_\br+\pi$ on any one of four sublattices of the
triangular lattice indicated earlier (we will work with these
`rotated' variables $\phi_\br$ and a ferromagnetic 4-spin term for
convenience).  This means shifting $\phi_\br \to \phi_\br+\pi$ on any
two such sublattices leaves the action invariant and is a symmetry
operation. We can identify this with a conservation of total boson
number modulo two on alternate rows of the triangular lattice. These
rows can run in any of the three symmetry directions of the lattice,
this corresponds to a total of four symmetry operations including the
identity.

The rotor Hamiltonian above describes bosons hopping on plaquettes of
the triangular lattice. For $J=0$, the dynamics conserves the center
of mass of the bosons, since the bosons hop equal distances in
opposite directions on any plaquette. In other words the center of
mass operator $\hat{\bR}_{\rm c.m.}=\sum_\br \br \hat{n}_\br$
satisfies $\left[\hat{\bR}_{\rm c.m.}, H\right]=0$, and is a constant
of motion.  Thus, in deriving the imaginary-time path integral for the
partition function in terms of $\phi$, we may insert factors of
$\exp(i\bq\cdot \hat{\bR}_{\rm c.m.})$ with an arbitrary vector $\bq$,
since $\hat{\bR}_{\rm c.m.}$ commutes with $H$. This corresponds to
$\phi_\br\to \phi_\br+\bq\cdot\br$ being a symmetry of the
phase-action.

If we identify $\phi=0$ with $\phi=2\pi$, or keep track of the fact
that the charge is quantized the first symmetry operation discussed
above is a special case of the second, and corresponds to choosing
$\bq=\bQ_i~(i=1,2,3)$ where the momenta $\bQ_i$ are at the center of
the edges of the Brillouin zone, indicated in Fig.~4(a).  The above
symmetries for $J=0$ imply we cannot have terms which depend on
$(\nabla \phi)$ in the phase action (in the path integral for the
partition function). The lowest allowed gradient term on which the
action can depend is of the form $(\nabla^2\phi)$.  We explicitly
confirm this in a spin-wave calculation in Section III.C, where we show
the collective mode disperses as $\omega^2(\bk) \sim |\bk|^4$ for
small $|\bk|$, and calculate its consequences for correlation
functions. The presence of a nonzero $J$ explicitly breaks these
symmetries, in fact $J=0$ will turn out to be a critical point of the
model.

\subsection{Plaquette duality and dual action}

For $J=0$, it is useful to define a dual representation for the model
which permits us to obtain spin-wave theory as a well-defined limit of
the model and to analyze instabilities of the spin-wave phase towards
charge and energy ordering. For this purpose, we use a plaquette
duality transformation \cite{bosemet} and work with dual variables $\theta_P$ 
and $N_P$ which reside at the centers of each of the 4-site plaquettes of
the triangular lattice --- these sites (labelled by $P$) form a Kagome
lattice with nearest neighbour spacing $1/2$ in units of the lattice
spacing of the original triangular lattice.  Alternatively, we may
label sites on the dual lattice as $(\br,\alpha)$, viewing the Kagome
lattice as a triangular lattice three sites (labelled $\alpha=1,2,3$)
per unit cell, and we will use this notation whenever
convenient. Thus, there are three times as many sites (and hence
degrees of freedom) on the dual lattice compared to the original, and
this is reflected in a redundancy in the description which we will
discuss shortly.
\begin{figure}
\begin{center}
\vskip-2mm
\hspace*{0mm}
\fig{2.4in}{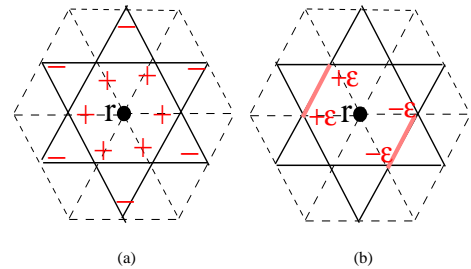}
\vskip-2mm
\caption{(a) The original triangular lattice (dashed lines) and the
  dual Kagome lattice (bold lines). The sites '+' and '-' indicate the
  sites of the Kagome unit around the site $\br$ of the triangular
  lattice, and are referred to as 'hex' and 'star' respectively in the
  paper. (b) The ``gauge'' transformation on the dual lattice ---
  shifting $\theta$ by $\varepsilon$ on the indicated sites as shown,
  with $\varepsilon/\pi$ being an integer, is an exact symmetry of the
  dual Hamiltonian.}
\label{fig:dual}
\end{center}
\end{figure}
Define \bea
\pi N_P&=&(\phi_1-\phi_2+\phi_3 -\phi_4) \label{d1}\\
\pi n_\br&=&\sum_{{\rm hex}} \theta_P - \sum_{{\rm star}} \theta_P
\label{d2}, \eea where the angles $\phi_1\ldots\phi_4$ in
Eq.(\ref{d1}) are around plaquette $P$ of the lattice with the
convention indicated in Fig.~1(a). The sites ${\rm hex}$ and ${\rm
  star}$ lie on the 12-site unit of the Kagome lattice around the site
$\br$ as shown in Fig.~2(a) (marked by '+' and '-' respectively),
while the center of each Kagome unit lies on the triangular lattice
(and is labelled $\br$). In these variables, the dual Hamiltonian on
the Kagome lattice is
\bea H_{\rm dual}&=&K \sum_P \cos(\pi N_P) \nn\\
&+& \frac{U}{2 \pi^2} \sum_\br (\sum_{{\rm hex}} \theta_P - \sum_{{\rm
    star}} \theta_P-\pi \bar{n})^2 \eea and the dual variables satisfy
$\big[N_P,\theta_{P'}\big]= i\delta_{P,P'}$, with $\theta_P/\pi$
having integer eigenvalues, while $N_P$ has a continuous spectrum.

We pause to elaborate further upon the redundancy of the $\theta$
variables.  Since only the combination $n_\br$ of the $\theta_P$
variables is physical, the Hamiltonian will be invariant under any
integer multiple of $\pi$ shift of the $\theta_P$ that leaves all the
$n_\br$ invariant.  This can be done locally as follows.  Pick any
hexagon of the dual Kagome lattice, and choose two neighboring sites
on this hexagon.  Shift $\theta_P \rightarrow \theta_P + \pi$ on those
sites, and simultaneously take $\theta_P \rightarrow \theta_P - \pi$
on the sites on the opposite side of this hexagon, leaving the
remaining two site of the hexagon (and all the other sites not on this
hexagon) untouched as shown in Fig.~2(b). 
Since this leaves all physical properties
invariant, this should be regarded as a local gauge symmetry.  Such
transformations are generated by the unitary operators
\begin{equation}
  \label{eq:gaugetransf}
  G_\br({\bf K}_\alpha) = e^{i {\bf K}_\alpha \cdot \sum_{\rm hex}
    ({\bf r}_P - \br) N_P},
\end{equation}
where ${\bf K}_\alpha$ with $\alpha=1,2,3$ are the three reciprocal
lattice vectors for the triangular lattice.  In the spirit of a gauge
theory, we could restrict our consideration to physical
gauge-invariant states which satisfy $G_\br({\bf K}_\alpha) =1$.  This
can be rewritten suggestively as
\begin{equation}
  \label{eq:vortexcom}
  \sum_{\rm hex} ({\bf r}_P - \br) N_P = 0 ({\rm mod}\; {\bf a}),
\end{equation}
where ${\bf a}$ is an arbitrary Bravais lattice vector for the
triangular lattice.  As in Refs.~\cite{leon},\cite{bosemet}, the
operator $N_P$ may be regarded as the local ``vortex number'' (modulo
$2$) on the site $P$ of the dual lattice.  Thus the condition of
Eq.~(\ref{eq:vortexcom}) requires that the local {\sl vortex center of
  mass} on each hexagonal plaquette of the Kagome lattice vanishes --
the ambiguity by a Bravais lattice vector corresponding exactly to the
ambiguity in $N_P$ by a shift of $2$.  Thus the gauge-invariance of
the dual description is related to the immobility of ``vortices'' in
this model.

Using a Trotter decomposition, we may write the partition function as
a discretized imaginary-time path integral in the standard manner.
This leads to $Z=\int {\cal D}\theta(\br,\tau)
\exp(-S\left[\theta(\br, \tau)\right])$ with the action \bea S_{\rm
  dual}&=& \frac{1}{\pi^2}\ln(\frac{2}{\epsilon_\tau K})
\sum_{\br,\alpha,\tau}
(\theta_{\br,\alpha,\tau}-\theta_{\br,\alpha,\tau+1})^2 \nn\\
&+& \frac{\epsilon_\tau U}{2\pi^2} \sum_\br(\sum_{\rm hex}
\theta_P-\sum_{\rm star} \theta_P-\pi\bar{n})^2 \eea where the field
$\theta_{\br,\alpha,\tau}/\pi$ is an integer-valued field,
$\epsilon_\tau=\beta/N_\tau$ with $\beta=1/T$ is the inverse
temperature, and $N_\tau$ is the number of imaginary time slices.  The
quantum problem at a temperature $T$ corresponds to the continuum
limit $\epsilon_\tau \to 0$, $N_\tau\to\infty$ with fixed
$1/\epsilon_\tau N_\tau = T$. Thus, we reduce the quantum model to an
anisotropic $(2+1)$-dimensional classical model --- since there is no
sign-problem, this proves useful for carrying out Monte Carlo
simulations in the dual representation to numerically test for charge
ordering and plaquette-energy ordering in the ground state at
arbitrary $U/K$.

\subsection{Effective theory and spin-wave approximation}
We may rewrite the dual action as
\bea
S_{\rm dual}&=&
\frac{1}{\pi^2}\ln(\frac{2}{\epsilon_\tau K}) \sum_{\br,\alpha,\tau}
(\theta_{\br,\alpha,\tau}-\theta_{\br,\alpha,\tau+1})^2 \nn\\
&+& \frac{\epsilon_\tau U}{2\pi^2}
\sum_\br(\sum_{\rm hex} \theta_P-\sum_{\rm star} \theta_P-\pi\bar{n})^2\nn\\
&-& \sum_{q,\br,\tau,\alpha} v^0_{2q} \cos(2 q \theta_{\br,\alpha,\tau})
\eea
where the bare couplings $v^0_{2q}$ are chosen to enforce 
the integer constraint on $\theta$, and may be singular. We will now
write down a low energy effective action in terms of
$\vartheta_{\br,\alpha,\tau}=[\theta]_f - (Q m n)/2$. Here
$[\theta]_f$ symbolically denotes an integration over the fast 
(high-energy) degrees 
of freedom while retaining the spatial structure of the lattice, 
$Q=2\pi \bar{n}$, and the site $\br\equiv m\hat{a}+n\hat{b}$ in terms
of the basis vectors $\hat{a},\hat{b}$ of the triangular lattice shown
in Fig.~1(b). The factor we have
subtracted out eliminates the mean density $\bar{n}$ from the
second (boson repulsion) term in the Hamiltonian. 

If we ignore the terms
with $v^0_{2q}$, we obtain a Gaussian action in terms of $\vartheta$.
Studying this action
motivates us to guess that the effective low energy action may
be of the form 
$S^{\rm eff}_{\rm dual}\!\!=\!\! S^{(0)}_{\rm dual}+S^{(1)}_{\rm dual}$
with
\be
S^{(0)}_{\rm dual}\!\!=\!\!
\int_ \omega\sum_{\bk,\alpha,\beta} \vartheta_{-\bk,\alpha,-\omega}
\left(\frac{\omega^2 \delta_{\alpha\beta}}{2 \pi^2 {\cal K}(\bk)}+
\frac{{\cal U}(\bk)}{2\pi^2} G_{\alpha\beta}(\bk) \right)
\vartheta_{\bk,\beta,\omega}
\ee
and
\be
S^{(1)}_{\rm dual}\!\!=\!\!
\int_\tau \sum_{q,\br,\alpha} v_{2q}
\cos(2q \vartheta_{\br,\alpha,\tau} + Q q m n)
+S_{\rm higher}
\ee
where we have defined renormalized couplings ${\cal U}(\bk), {\cal K}(\bk)$
(which are smooth nonzero functions of $\bk$)
and $v_{2q}$, the site $\br\equiv(m,n)$ as before, and
the $3\times 3$ matrix $G_{\alpha,\beta}(\bk) =
A^\ast_\alpha (\bk) A_\beta(\bk)$ with
\bea
A_1(\bk)&=&1+e^{-i k_a}-e^{-i k_a -i k_b} - e^{i k_b} \nn\\
A_2(\bk)&=&1+e^{-i k_a -i k_b}-e^{-i k_a} - e^{- i k_b}\nn\\
A_3(\bk)&=&1+e^{-i k_b}-e^{i k_a} - e^{- i k_a -i k_b},
\eea
where $k_a=\bk\cdot\hat{a}$ and $k_b=\bk\cdot\hat{b}$.
$S_{\rm higher}$ denotes other terms which might be generated 
in deriving the low energy action. For the present 
purposes, we will not write down the explicit form of these terms
but we will discuss them later.

If we set $v_{2q}=0$ and ignore $S_{\rm higher}$, we can obtain
the eigenmodes of the Gaussian action $S^{(0)}_{\rm dual}$
by diagonalizing $G_{\alpha\beta}(\bk)$. We
find there is one nonzero eigenvalue $\lambda(\bk)$, and two zero 
eigenvalues. This corresponds to one 
dispersing mode ($\Theta_1$) and two nondispersive 
modes ($\Theta_{2,3}$)
with zero energy (flat bands) and the action at this level 
takes the form
\bea
S^{(0)}_{\rm dual}&\!\!=\!\!&
\int_ \omega\sum_{\bk}
\left(\frac{\omega^2+{\cal E}^2(\bk)}{2\pi^2 {\cal K}(\bk)}\right)
|\Theta_1(\bk,\omega)|^2 \nn \\
&+& \int_ \omega\sum_{\bk,\alpha=2,3}
\left(\frac{\omega^2}{2 \pi^2 {\cal K}(\bk)}\right)
|\Theta_\alpha(\bk,\omega)|^2,
\eea
where we have set $\lambda(\bk)={\cal E}^2(\bk)/{\cal U}(\bk){\cal K}(\bk)$
so that ${\cal E}(\bk)$ appears as the excitation energy of the collective
mode $\Theta_1$. Explicitly,
\bea
{\lambda}(\bk)&=&2 [ 6-3\cos k_a - 3\cos k_b
-3\cos(k_a+k_b) \\
&+&\cos(k_a-k_b) +\cos(2 k_a + k_b) +\cos(2 k_b + k_a)] \nn,
\eea
thus the dispersing mode $\Theta_1$ has one gapless point at
$\bk=(0,0)$. The leading dispersion away from this point is
${\cal E}^2(\bk \to 0) = (9/16){\cal U}(0){\cal K}(0) |\bk|^4$ ---
as stated in subsection A, this dispersion is dictated by the
symmetry of conservation of center of mass of the bosons which
forbids a term $\sim |\bk|^2$.

We can also now see that the extra degrees of freedom (two per site of
the triangular lattice) introduced in going over to the dual
description manifest themselves as zero energy modes corresponding to
unphysical fluctuations, while the dispersing mode corresponds to
physical fluctuations.  In fact, by Fourier transforming a general
superposition $f_2(\bk)\Theta_2(\bk) + f_3(\bk) \Theta_3(\bk)$, we can
show that the zero modes indeed arises from the local gauge symmetry
generated by Eq.~(\ref{eq:gaugetransf}) in real space.  More
precisely, in the coarse-grained quadratic low energy theory, these
discrete symmetry operations are promoted to continuous shifts (with
arbitrary $|{\bf K}_\alpha|$).  Similarly, Fourier transforming
$f_1(\bk) \Theta_1(\bk)$, we find that the physical fluctuations are
composed precisely of the gauge invariant boson densities.

At this stage, we may reintroduce the boson phase fields by a Hubbard
Stratonovitch decoupling of the time derivative term of the physical 
fluctuation $\Theta_1$ (as in Ref.~\cite{bosemet}), namely
\bea
e^{-\frac{\omega^2}{2\pi^2 {\cal K}(\bk)} 
|\Theta_1(\bk,\omega)|^2}
&=& \int D\phi\phi^*
e^{-\frac{{\cal E}^2(\bk)}{2{\cal U}(\bk)} 
|\phi(\bk,\omega)|^2 }\\
&\times& 
e^{\frac{\omega{\cal E}(\bk)}{2\pi\sqrt{{\cal U}(\bk){\cal K}(\bk)}}
\left(\phi(\bk) \Theta^*_1(\bk) -
\phi^*(\bk) \Theta_1(\bk)\right)}.\nn
\eea
Provided $v_{2q}$ and $S_{\rm higher}$ are irrelevant and
can be ignored, we can integrating out the field $\Theta_1$ 
which leads to a Gaussian action for the boson field $\phi$,
\be
S^{(0)}_{\phi}=
\int_\omega \sum_\bk 
\left(\frac{\omega^2 + {\cal E}^2(\bk)}{2{\cal U}(\bk)}\right)
|\phi(\bk)|^2.
\ee
This corresponds to a spin-wave (harmonic) approximation 
and may be used to evaluate boson correlation functions
to zeroth order in $v_{2q}$.

\subsection{Boson correlation functions in spin-wave theory}
To characterize the Gaussian theory $S^{(0)}_\phi$, we 
evaluate the space-time correlation functions for the boson
creation operator $\exp(i\phi_{\br\tau})$. At long times or
large separations, these correlation functions reduce to
\bea
\la e^{i\phi_{\br,0}} e^{-i\phi_{\bo,0}} \ra &\sim& |\br|^{-\frac{1}{2\pi
\sqrt{3}} \sqrt{\frac{{\cal U}(0)}{{\cal K}(0)}}} \nn\\
\la e^{i\phi_{\bo,\tau}} e^{-i\phi_{\bo,0}} \ra &\sim& |\tau|^{-\frac{1}{4\pi
\sqrt{3}} \sqrt{\frac{{\cal U}(0)}{{\cal K}(0)}}}.
\eea
Clearly, the Gaussian theory describes a critical liquid of the
bosons with power law correlations arising from the gapless
excitations dispersing as ${\cal E}(\bk) \sim |\bk|^2$. 
From the long time behavior of the two-point correlation in a 
finite size system, we may evaluate the finite size scaling of the 
gap to adding a particle to be just $\Delta(L) \simeq {\cal U}(0)/L^2$.
In the thermodynamic system, the low energy density of states per unit 
volume of collective excitations is given by
\bea
N(\omega)&=&\frac{1}{V} \sum_\bk \delta(\omega-{\cal E}(\bk))\nn\\
&\stackrel{\simeq}{_{\omega\to 0}}
& \frac{1}{2\pi\sqrt{3 {\cal U}(0) {\cal K}(0)}}
\eea
which is a constant depending on the interactions.
Thus, all
low-energy long-wavelength properties of the liquid are determined
in terms of the $\bk \to 0$ behavior of the functions 
${\cal U}(\bk), {\cal K}(\bk)$. We next present arguments and
numerical evidence that this Gaussian description of a
critical liquid in the model with $J=0$ may be valid even in
the presence of terms $v_{2q}$ and $S_{\rm higher}$.

\subsection{Argument for irrelevance of higher order terms}
To see whether the term $v_{2q}$ is relevant, we can evaluate
the correlation functions of this operator, and we find it is
local in space and exponentially decaying in time,
\be
\la \cos 2\vartheta_{\br,\alpha,\tau} \cos 2\vartheta_{\bo,\beta,0} \ra
\sim \delta_{\br,0} \delta_{\alpha,\beta}~\exp(-\gamma |\tau|)
\ee
with a constant $\gamma > 0$. This is not 
hard to understand --- it arises from the fact that the variable
$\vartheta$ is a combination of the eigenmodes $\Theta_{1,2,3}$
and $\cos 2\vartheta_{\br,\alpha,\tau}$ is not ``gauge-invariant''.
As a result its spatial correlations are local, while
the fluctuations of the zero-modes $\Theta_{2,3}$ determine
the exponential temporal decay. Clearly, $v_{2q}$ is irrelevant.

Consider the possible forms we can obtain for $S_{\rm higher}$.
These terms would be of the form of cosines involving
multiple $\vartheta$'s at different space-time points. However,
similar arguments as above apply to these operators. The only operators
which can have nonzero correlations at nonzero separation would
be gauge-invariant combinations of the $\vartheta$'s. As we have
shown, these are the local boson densities. Thus, if we admit only
slow fluctuations of $\Theta_1$, these would take the form of weak
density-density interactions within a perturbative treatment and
could renormalize the coefficients of the Gaussian action, but not
cause an instability. Of course, such density-density interactions
might lead to a charge-ordered state at strong couplings, but our
arguments show that the Gaussian theory is perturbatively stable
\cite{footnote}.

\subsection{Numerical results}
\begin{figure}
\begin{center}
\vskip-2mm
\hspace*{0mm}
\fig{3.0in}{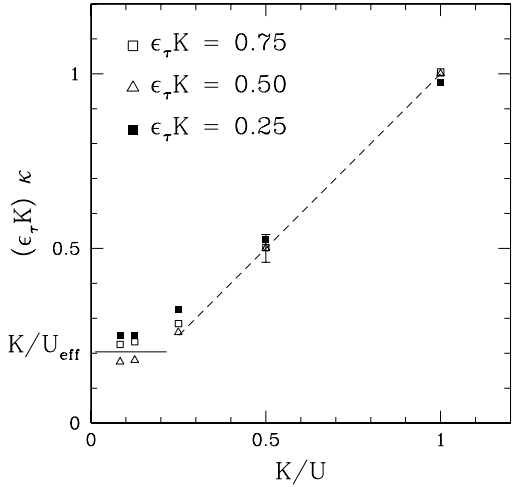}
\vskip-2mm
\caption{The dependence of the scaled compressibility $(\epsilon_\tau K) 
\kappa$ 
on the bare boson repulsion $U$, for various $\epsilon_\tau K$. 
Within spin-wave theory, we expect 
$(\epsilon_\tau K) \kappa = K/{\cal U}(0)$ as discussed in the text. From 
the figure we see that $K/{\cal U}(0)$ coincides with $K/U$ for small $U$ 
(as indicated by the dashed line), but
deviates and tends to a constant $K/U_{\rm eff} \sim 0.2$ as the bare $U \to
\infty$ (as shown by the solid line drawn as a guide to the eye). 
Typical error bars are indicated at one point. We thus 
expect the $S=1/2$ $XY$ spin model with $J=0$ to
be compressible, with ${\cal U}(0)=U_{\rm eff}$.}
\label{fig:chi}
\end{center}
\end{figure}
To confirm the above arguments we have carried out Monte Carlo
calculations on the dual model using a Metropolis algorithm, for
lattice sizes upto 432 spatial-sites and 48 time-slices, and
periodic boundary conditions. We find no evidence of any charge
ordering for $\bar{n}=1/2$, even for large $U/K$. We have
compared the density-density correlations and find agreement
with what we expect from a Gaussian theory $S^{(0)}_{\rm dual}$.
In particular, we show results for the compressibility $\kappa$
defined through $\kappa=\chi_{nn}(\bk\to 0,\omega_n=0)$ where
\be
\chi_{nn}(\bk,\omega_n)= \sum_{\br,\tau} e^{-i\bk\cdot\br+
i\omega_n\tau} \la n(\br,\tau) n(\bo,\tau) \ra.
\ee
Within the Gaussian theory, $\kappa=1/\epsilon_\tau {\cal U}(0)$.
In Fig.~3 we plot the scaled compressibility $(\epsilon_\tau K)\kappa$
which is expected to scale as $K/{\cal U}(0)$. We find that
that it is nonzero at all $U/K$ in the quantum ($\tau$-continuum) 
limit, with ${\cal U}(0)/K \simeq 5$ as $U/K \to \infty$. We have
also checked and found no evidence for energy ordering on the plaquettes
of the triangular lattice. Thus, we believe that for $\bar{n}=1/2$,
the ring-exchange model with $J=0$ is well described by
the Gaussian fixed-point action $S^{(0)}_{\rm dual}$
where terms with $v_{2q}$ and
all other terms $S_{\rm higher}$ are irrelevant.

\section{Perturbing in nearest neighbor coupling $J$}

\subsection{Ordered ground state}
To perturb in the nearest-neighbor exchange $J$, it is convenient to
work in the phase representation which we have argued remains a valid
`fixed-point' description of the pure ring-exchange model for all
values of the bare coupling $U/K$. In this, we make the approximation
of completely ignoring irrelevant operators --- this seems reasonable
in retrospect since we will find that the system acquires long-range
phase order at $T=0$.

The nearest-neighbor perturbation we add to the Gaussian phase action 
takes the form
\bea
S_J&=&J\sum_{m,n}\int d\tau\left[ (-1)^n \cos(\phi_{m,n,\tau}-\phi_{m+1,n,\tau})
\right.\nn\\
&+& (-1)^m \cos(\phi_{m,n,\tau}-\phi_{m,n+1,\tau}) \nn\\
&+& \left. (-1)^{m+n} \cos(\phi_{m,n,\tau}-\phi_{m+1,n+1,\tau}) \right]
\eea
where we have chosen to label sites on the triangular lattice as
$\br=m\hat{a}+n\hat{b}$, and we are working in rotated variables 
$\phi_\br$ for which the ring-exchange term is ferromagnetic.

To carry out a systematic expansion, we define
$\phi_\br=\tilde{\phi}_{\br}+\bQ\cdot\br$ with an arbitrary vector
$\bQ$. This allows us to examine at the effect of perturbing in $J$ in
any of the ground states of the degenerate manifold to see if
fluctuations select any state. Perturbing in $S_J$ around the pure 
ring-exchange Gaussian theory, we find that the corrections to the 
free energy vanish at leading order in $J$.
At ${\cal O}(J^2)$, we find a correction to the free energy density
\bea
\!\!&&\!\!\delta f=-\frac{1}{2} \beta J^2 \left(\cos^2 Q_a
+\cos^2 Q_b +\cos^2(Q_a+Q_b) \right) \nn \\
& &\!\!\!\!\!\!\times\sum_{m,n}\int_\tau (-1)^n \la 
\cos(\tilde{\phi}_{0,0,0}-\tilde{\phi}_{1,0,0}
+\tilde{\phi}_{m,n,\tau}+\tilde{\phi}_{m+1,n,\tau})\ra_0 \nn \\
\eea
where $\la\ldots\ra_0$ is the expectation value evaluated in the
Gaussian theory, $Q_a\equiv\bQ\cdot\hat{a}$, $Q_b\equiv\bQ\cdot\hat{b}$,
and we have used the six-fold symmetry of the triangular lattice to
simplify certain intermediate expressions. The expectation value may
be analytically evaluated at small $U/K$ by expanding the cosine term
since its argument is small in the Gaussian theory, and we find that
the sum over $m,n$ is positive.
Thus, (i) the corrections to the free energy are independent
of the sign of $J$, and (ii) minimizing $\delta f$ with respect to
variations of $\bQ$ corresponds to maximizing $\left(\cos^2 Q_a
+\cos^2 Q_b +\cos^2 (Q_a + Q_b) \right)$ which leads to ordering at
four wavevectors $\bQ_\alpha$ ($\alpha=1\ldots 4$) which are shown in
Fig.~4(a). We expect this order to persist even at large effective $(U/K)$
(recall that the bare $U/K \to \infty$ in the $S=1/2$ spin limit).
To obtain the ordered states in the original phase variables,
we rotate back $\phi(\br)\to\varphi(\br)=\phi(\br)-\pi$ on the triangular
sublattice and we then find, as shown in Fig.~4(b), that the 
resulting ordered states break 
(i) global $U(1)$ spin rotational invariance and (ii) four-fold 
translational invariance. For small $J/K$, 
such states have been shown to 
occur at finite magnetic fields for the classical version of a related 
SU(2) invariant ring-exchange model \cite{momoi}, and this 
ordering has been called $uuud$ (for `{\em u}p-{\em u}p-{\em u}p-{\em 
d}own' which is the spin order on a 4-site plaquette, but in the
plane in our case). Here, we find such
an ordered state is stabilized by quantum fluctuations in the $XY$
limit even in zero magnetic field. 

\subsection{Fluctuations in the ordered state}

As shown above, in the presence of a nonzero $J$, the model has an
ordered ground state. This order breaks U(1) invariance.
At the same time, it also breaks translational symmetry. 
In the boson language, this state may be viewed as a superfluid 
coexisting with broken translational order. This broken
discrete symmetry is also evident from our implicit choice of a state
$\bQ_\alpha$ to minimize the free energy. Thus, we expect two kinds of
excitations at low energies about the ordered state --- (i) phase 
fluctuations, and (ii) domain walls in the discrete order.

It is easy to see that with $\phi_{\br}=\bQ\cdot\br+\tilde{\phi}_\br$,
there is an energy cost $\sim (\bQ-\bQ_\alpha)^2$ for small 
deviations $|(\bQ-\bQ_\alpha)| \ll 1$ where $\bQ_\alpha$ denotes any
one of the ordering vectors. This implies that the long
wavelength effective theory of phase fluctations around each of these 
ordered states is described by an action of the form
\be
S = \int_{\br,\tau}\left( \frac{1}{U_{\rm eff}} (\partial_\tau \varphi)^2 
+ J_s (\nabla\varphi)^2 + K_{\rm eff} (\nabla^2\varphi)^2 \right)
\ee
with a nonzero $J_s$, and effective couplings
$U_{\rm eff}$ and $K_{\rm eff}$. For small $U/K, J/K$, we find 
$U_{\rm eff} \sim U$, $K_{\rm eff} \sim K$ and $J_s \sim J^2/K 
\sqrt{U/K}$. Thus, phase fluctuations are controlled by the phase
stiffness $J_s$.

A sharp domain wall separating regions with different $\bQ_\alpha$ 
order would cost an energy $\sim K$ per unit length. However, 
since the model is spin-disordered for $J=0$, we expect that a smooth
deformation of the spin configuration by making a domain wall with
nonzero width $\xi_D$, would cost less energy. Indeed, using the above
action one can estimate the energy per unit length of a straight 
domain wall as $e_{\rm dom} = \gamma_1 K_{\rm eff}/\xi_D +\gamma_2 
J_s \xi_D$. where $\gamma_{1,2} \sim 1$ are constants. Minimizing the 
energy cost with respect to $\xi_D$, we find that $\xi^\ast_D \sim 
\sqrt{K_{\rm eff}/J_s}$, and thus domain wall excitations cost
an energy $e^\ast_{\rm dom} \sim \sqrt{K_{\rm eff} J_s}$ per unit 
length. For small $U/K$, $e^\ast_{\rm dom} \sim J (U/K)^{1/4}$.

\subsection{Phase transitions at finite temperature}

At high temperature, we expect the model to be in a fully disordered
phase where the $U(1)$ symmetry as well as the discrete broken symmetry
is restored. To study these phase transitions, we appeal to the above
estimates of the spin stiffness and domain wall energies. For $U/K
\sim 1$, since the
spin stiffness $J_s \sim J^2/K$, while the domain wall energy per unit
length $e_{\rm dom} \gtrsim J \gg J_s$, we expect the $U(1)$ symmetry to be 
restored via a BKT transition \cite{bkt} at a temperature $T_{\rm BKT} 
\sim J_s$ once vortices are included in the effective theory, while the 
discrete symmetry is not restored until a much
higher temperature $T_c \sim J$. Thus there is an intermediate phase
with exponentially decaying spin correlations, but with the discrete
order still present. This bears resemblance to proposed scenarios 
of chiral ordering in the $XY$ antiferromagnet on a triangular
lattice as well as the $SU(2)$ multiple-spin exchange model. However,
in contrast to the $XY$ model where a single scale, namely the two-spin
exchange coupling, sets the scale for both transition temperatures,
the transition temperatures are governed by distinct energy scales
in the present case, and the discrete symmetry corresponds to
broken translational invariance.

\begin{figure}
\begin{center}
\vskip-2mm
\hspace*{0mm}
\fig{3.0in}{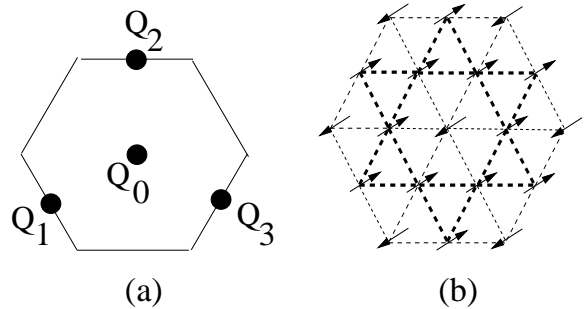}
\vskip-2mm
\caption{(a) The four ordering wavevectors $\bQ_\alpha$ ($\alpha=1\ldots
4$) obtained by perturbing in the nearest neighbor coupling $J$, with the
hexagon indicating th first Brillouin zone of the triangular lattice. (b) 
The phase/spin order in the original variables $\varphi_\br$ in 
one of the
four broken symmetry ground states. The order may be most easily thought of
as alternating rows of ferromagnetic and antiferromagnetic spins. The bold
dashed lines indicate bonds on which the spins point in the same direction,
while the other bonds have antiferromagnetic spin orientation. As discussed
in the text, these correspond to energy ordering in the ground state with
an associated four-fold broken discrete symmetry.}
\label{fig:groundstate}
\end{center}
\end{figure}

In order to better understand the discrete transition, let us identify
the appropriate order parameter and construct a Landau theory. Since
the discrete symmetry is easily identified with translations in the
original variables $\varphi_\br$, we will discuss it in terms of that.  
As shown in Fig.~4(b), the ordered state at $T=0$ has broken translational
invariance associated with ordering of $\cos (\varphi_i-\varphi_j)$
on nearest-neighbor bonds. Since there is a nearest-neighbor spin
exchange term $J$, this leads to ordering in the energy density. We
identify the broken discrete symmetry with energy ordering on the bonds. 
To identify the appropriate order parameter for this transition, note
that we may write $K_{\br,i}=\la \bS_\br \cdot \bS_{\br+\hat{a}_i}\ra$
where $\hat{a}_i~(i=1,2,3)$ are the unit lattice vectors making an
angle $2\pi/3$ with each other (specifically, $\hat{a}_1 = \hat{a}$,
$\hat{a}_2 = \hat{b}$, and $\hat{a}_3 = -(\hat{a}+\hat{b})$). Labelling
the four ordered states by $\mu=0\ldots 4$ and with $a_0=0$, we can write
the expectation value in any of these ordered states as
$K^\mu_{\br,i}=\cos(\bQ_i\cdot\br) \cos(\bQ_i\cdot \hat{a}_\mu)$. For
a general superposition,
\be
K_{\br,i}=\sum_{\mu=0\ldots 4} 
B_\mu \cos(\bQ_i\cdot\br) \cos(\bQ_i\cdot \hat{a}_\mu)
\ee
and we identify $B_\mu$ as the appropriate order parameter 
for the Landau theory. Under shifts of $B_\mu \to B_\mu + \lambda$,
the physical correlation $K_{\br,i}$ is unchanged. Thus we
require the Landau functional to be invariant under such shifts.
Further, studying the transformation of $K_{\br,i}$ under symmetry
operations of the lattice (unit translations, $\pi/3$-rotations,
and reflections about the $3$ mirror planes), we find that such
symmetry operations correspond to all possible permutations of
the set $\{B_\mu\}$. For the Landau theory we find beyond quadratic
order, one cubic invariant and two quartic invariants. It appears 
likely that the finite temperature phase transition restoring this
discrete symmetry is in the same universality class as a $4$-state 
Potts model.
%\cite{zia}.

To summarize, we expect the system to exhibit with increasing 
temperature a BKT transition with $T_{\rm BKT} \sim J^2/K$ from a state 
with power law spin 
correlations to a state with exponentially decaying spin correlations.
The discrete broken symmetry associated with energy ordering on
the bonds of the lattice is expected to be restored above a temperature
$T_c \sim J$, with the transition being in the universality class
of the 4-state Potts model. In a weak applied in-plane magnetic field, 
the $U(1)$ invariance is lost, and the BKT transition would become
rounded but the discrete transition would survive as a finite temperature
transition. At large enough fields, we expect a first order transition
to a state where all the spins point in the same direction and
the translation invariance is restored.

\section{discussion and conclusions}

Studying an $XY$ model with four-spin couplings and small
two-spin exchange, we have obtained a ferrimagnetically ordered
ground state and analyzed phase transitions out of this phase.
Since the ground state is phase ordered, we expect our result to
be stable to small out-of-plane couplings involving $S_z$. 
Similarly, introducing weak couplings between such ordered 
two-dimensional planes would lead to a three-dimensionally
ordered state \cite{tsuneto}, 
where the spin/phase order would persist to finite
temperature. We thus expect this state might be
of some interest for Mott insulators on a triangular lattice.
If such an ordered state exists, it could be detected
in neutron diffraction studies. The spin-disordered state
with persisting discrete broken symmetry may be indirectly observable
through lattice distortions if the spins couple to the lattice.

We can also analyze the opposite limit from the one studied in
the paper, where we assume ``easy
axis'' anisotropy in the model in Eq.~(\ref{hubbard}) and weak
$J/K$. In this Ising limit, we find eight degenerate
ground states for the spin model, again of the $uuud$ type, with the 
spins aligned along $S_z$-axis. These
states correspond to charge-ordered incompressible states for
the bosons at density $\bar{n}=1/4$ or $\bar{n}=3/4$ (four ground
states at either density). It may be 
that any anisotropy which takes us away
from the SU(2) invariant model leads to ordered ground states, while the 
$SU(2)$-invariant model seems to be a uniform spin liquid 
\cite{lhuillier,imada}.
It would be interesting to further explore this possibility.

\begin{acknowledgments}
  This work was supported by the NSF through grants DMR-9985255 and 
  PHY99-07949, and by
  the Sloan and Packard foundations.  We would like to thank Luca
  Capriotti, Matthew
  Fisher, Eduardo Fradkin, Akakii Melikidze and Ashvin Vishwanath for useful
  discussions and comments on the manuscript.
\end{acknowledgments}


\begin{thebibliography}{999}
\bibitem{thouless} 
D.J. Thouless, Proc. Phys. Soc. Lond. {\bf 86}, 893 (1965).
\bibitem{hetherington} 
J.M. Delrieu and J.H. Hetherington, Rev. Mod. Phys. {\bf 55}, 1 (1983).
\bibitem{ceperley}
B. Bernu, L. Candido and D.M. Ceperley, Phys. Rev. Lett. {\bf 86},
870 (2001).
\bibitem{chakravarty}  
K. Voelker and S. Chakravarty, Phys. Rev. B {\bf 64}, 235125 (2001).
\bibitem{macdonald}  
A.H. Macdonald, S.M. Girvin and D. Yoshioka,
Phys. Rev. B {\bf 37}, 9753 (1988).
\bibitem{coldea}
R. Coldea, S.M. Hayden, G. Aeppli, T.G. Perring, C.D. Frost, T.E. 
Mason, S.-W. Cheong, and Z. Fisk,
\bibitem{lhuillier} G. Misguich, B. Bernu and C. Waldtmann,
Phys. Rev. B{\bf 60}, 1064 (1999); W. LiMing, P. Sindzingre and G. Misguich,
Phys. Rev. B{\bf 62}, 6372 (2000).
\bibitem{leon} L. Balents, S.M. Girvin and M.P.A. Fisher, 
Phys. Rev. B {\bf 65}, 224412 (2002).
\bibitem{motrunich}
T. Senthil and O. Motrunich, cond-mat/0201320;
O. Motrunich and T. Senthil, cond-mat/0205170;
O. Motrunich, cond-mat/0210684.
\bibitem{moessner}
R. Moessner and S. Sondhi, Phys. Rev. Lett. {\bf 86},
1881 (2001).
\bibitem{hermele} M. Hermele, unpublished (2002).
\bibitem{luca}
See for e.g., L. Capriotti, R. Vaia, A. Cuccoli and V. Tognetti,
Phys. Rev. B {\bf 58}, 273 (1998), and references therein.
\bibitem{bosemet}
A. Paramekanti, L. Balents and M.P.A. Fisher, 
Phys. Rev. B {\bf 66}, 054526 (2002).
\bibitem{bkt}
V.L. Berezinskii, Zh. Eksp. Teor. Fiz. {\bf 59}, 907 (1970)
[Sov. Phys. JETP {\bf 32}, 493 (1971)];
J.M. Kosterlitz and D.J. Thouless, J. Phys. C {\bf 6}, 1181 (1973);
J.M. Kosterlitz, J. Phys. C {\bf 7}, 1046 (1974).
\bibitem{imada}
H. Morita, S. Watanabe and M. Imada, cond-mat/0203020;
S. Watanabe and M. Imada, cond-mat/0207550.
\bibitem{footnote}
The fixed point boson action for $J=0$ resembles those studied 
in Ref.~\cite{lifshitz}, in connection with classical three-dimensional 
smectics, which exhibit a line of fixed points. However, since our 
starting point is a
microscopic $S=1/2$ quantum spin model, the phases and phase transitions 
in our model, as well as the duality mapping we use, are all quite 
different.
\bibitem{momoi}
T. Momoi, K. Kubo and K. Niki, Phys.
Rev. Lett. {\bf 79}, 2081 (1997);
K. Kubo and T. Momoi, Z. Phys. B {\bf 103}, 485 (1997);
T. Momoi, H. Sakamoto and K. Kubo,
Phys. Rev. B {\bf 59}, 9491 (1999).
%\bibitem{zia}
%For a field-theoretic formulation of the Potts model, see 
%R.K.P. Zia and D.J. Wallace, J. Phys. A: Math. Gen. {\bf 8}, 1495 
%(1975).
\bibitem{tsuneto}
S. Hikami and T. Tsuneto, Prog. Theor. Phys. {\bf 63}, 387 (1980).
\bibitem{lifshitz}
G. Grinstein, Phys. Rev. B {\bf 23}, 4615 (1981); D.J. Amit, S. Elitzur,
E. Rabinovici and R. Savit, Nucl. Phys. B {\bf 210}, 69 (1982).
\end{thebibliography}
\end{document}